# Fabrication and characterization of the gapless half-Heusler YPtSb thin films

Wenhong Wang,[1,a)] Yin Du,[1] Enke Liu,[1] Zhongyuan Liu,[2] and G. H. Wu[1]

[1]*Beijing National Laboratory for Condensed Matter Physics, Institute of Physics, Chinese Academy of Sciences, Beijing 100080, P. R. China*

[2]*State Key Laboratory of Metastable Material Sciences and Technology, Yanshan University Technology, Qinhuangdao 066004, P. R. China*

Half-Heusler YPtSb thin films were fabricated by magnetron co-sputtering method on MgO-buffered $SiO_2$/Si(001) substrates. X-ray diffraction pattern and Energy dispersive X-ray spectroscopy confirmed the high-quality growth and stoichiometry. The temperature dependence of the resistivity shows a semiconducting-type behavior down to low temperature. The Hall mobility was determined to be 450 $cm^2$/Vs at 300K, which is much higher than the bulk value (~300 $cm^2$/Vs). In-plane magnetoresistance (MR) measurements with fields applied along and perpendicular to the current direction show opposite MR signs, which suggests the possible existence of the topological surface states.

Topological insulators (TIs) are of a new class of materials, which has a bulk band gap generated by strong spin-orbit coupling but contains gapless surface states.[1-3] Uniquely these states are robust with respect to disorder and exhibit high mobility due to the suppression of carrier backscattering. In addition, theses sates are spin-momentum locked, making TIs attractive for future spintronic devices and quantum computing applications as well.[4,5] Since the first theoretical prediction of a two-dimensional TI in an HgTe based quantum well,[6,7] several other families of materials for three-dimensional (3D) TIs have been proposed theoretically and well-studied experimentally.[8-16] For example, tetradymite semiconcuctors, such as $Bi_2Te_3$, $Bi_2Se_3$, and $Sb_2Te_3$ are confirmed to be 3D TI with a single Dirac-cone on the surface, where the bulk band gap is as large as 0.3 eV, making the room temperature application possible.[8-12] However, a clear shortcoming of $Bi_2Te_3$ family is that these materials cannot be made with coexisting magnetism, a much desired property for spintronic applications. Although doping can be used to achieve the magnetically ordered behavior,[17] this creates extra complexity in material growth and could introduce detrimental effects upon doping.

Recently, half-Heusler ternary compounds with 18 valence electrons has been predicted to be 3D TI by alloying or proper stain engineering.[18-21] Most importantly, it was proposed that in the half-Heusler family the topological insulator allows the incorporation of superconductivity and/or magnetism. Among the experimentally studied TI candidates of half-Heusler materials, the gapless YPtSb is found to exhibit very good thermoelectric properties with a high figure of merit ZT of 0.2 and the high

Hall mobility, and thus has become the potential material of choice for both experimental and theoretical studies. [22, 23] Development of devices utilizing TIs will require accurate characterization of the electronic transport. However, no transport investigation on the half-Heusler YPtSb thin films was carried out so far. The aim of this work is to try the way to obtain high quality YPtSb thin films, and to investigate their structure and transport properties.

The YPtSb films were deposited by co-sputtering of three facing magnetron guns equipped with Y, Pt and Sb targets, respectively. This technique allows control of the composition of the deposited film by the relative Y, Pt and Sb deposition rates. The diameter of the target is 54.8 mm, and the purity is 99.9% for Y and Pt, while 99.9999% for Sb. The YPtSb film was deposited on cleaned $SiO_2$/Si(001) substrate with a 10-nm-thick MgO buffered. The MgO buffer layer was deposited by rf sputtering directly from a sintered MgO target under an Ar pressure of 10 mTorr. The YPtSb films were deposited at $550^0$C substrate temperature in a pure Ar gas condition and at a system pressure of 0.1Pa. The deposition rate of each target was first calibrated and the growth rate of the co-sputtered YPtSb films was 0.035 nm/s. Moreover, to avoid the surface oxidization, the YPtSb films were capped with a 2nm thick MgO for the structure and transport characterization.

The X-ray diffraction (XRD) pattern of YPtSb film is shown in Fig. 1 (a), and all the diffraction peaks can be assigned to the half-Heusler phase with the $C1_b$ type structure (F43m space group). By analyzing and comparing our XRD pattern with standard one, we found the YPtSb film has a slightly (111)-preferred orientation along the thickness direction. The lattice parameter was determined to be 6.53Å for YPtSb film which is in good agreement with the bulk value reported in ref. 12. Energy dispersive X-ray spectroscopy (EDX) measurements were carried out to check the stoichiometry for the YPtSb thin films. The results of quantitative composition analysis of the two typical YPtSb films by EDX are summarized in Table 1. It showed that the Y, Pt and Sb mole fraction ratio is close to 1:1:1. The measurement uncertainty of less 1.0 % was determined by repeating measurements on the same sample surface. The topography of the YPtSb thin film was observed using field emission scanning microscopy. A fine-grained structure with average grain sizes of 40 nm and an average roughness of 0.5 nm were observed (Fig. 1(b)). Each of the grains perhaps contained a few crystalline particles with mean sizes of 25 nm as estimated by the Scherrer method from XRD data.

Figure 2(a) shows the electrical resistivity $\rho_{xx}(T)$ in the range from 2K to 300 K. The increase in resistivity with decreasing temperature is a clear signature that the studied YPtSb film exhibits clearly a semiconducting-type behavior. The resistivity of the sample increases slowly from 300 K to 30 K. At temperatures below 30 K, however, the resistivity increases quickly with decreasing temperature, which is consistent with the recent report by Ouardi et. al.[22] in the bulk YPtSb polycrystalline sample. The inset of Figure 2(a), we show the temperature-dependent magnetic susceptibility. It exhibits no sign of magnetic order, indicating a time-reversal-invariant spin-orbit (diamagnetic) ground state in the YPtSb thin film, which is a necessary condition for realizing a topological insulator state.

The carrier concentration $n_e$ and mobility $\mu_e$ of YPtSb film were determined by measurements of the Hall effect. The Hall resistance of the YPtSb film was measured in the temperature range from 5K to 300K in magnetic induction fields from +5T to −5 T. The Hall coefficient was calculated from the slope of $R_H$ curves. The Hall mobility $\mu_e=R_H/\rho$ and the carrier concentration $n_e=1/eR_H$ were extracted from RH using a single band model. As shown in Figure 3(b), the Hall mobility $\mu_e$ of PtYSb increases first and then keep almost unchanged with temperatures up to 300K. We found that the $\mu_e$ reaches a constant value of ~450 cm$^2$/Vs at 300 K, which is higher than the bulk value of YPtSb (~300 cm$^2$/Vs).[22] Here we should point out that the $\mu_e$ strongly depends on thickness for thin films but becomes independent of film thickness beyond 5 μm approaching its bulk value. The enhanced mobility is therefore attributed to overall scattering due to grain boundaries and surface effects. At low temperatures below 50 K, the carrier concentration has a constant value of about $n_e \approx 5.0\times10^{18}$ cm$^{-3}$. Above 50 K, the carrier concentration increases almost linearly. The rate of increase of $n_e$ with increasing temperature becomes higher with film thickness. Therefore, the low value of $n_e$ (16×10$^{18}$ cm$^{-3}$ at 300 K) strongly indicate that the YPtSb film is a gapless semiconductor as revealed by the spectroscopic investigation in Ref. 22.

The magnetoresistance (MR) properties of YPtSb film under an in-plane field magnetic field are investigated as a function of the angle between the current and the magnetic field. The results are shown for two different conditions as depicted in Figure 3. A constant current of 5 mA was placed in the film plane, and a magnetic field up to 8T was applied to the pane in two configurations: parallel (a) and perpendicular (b) to the current direction, respectively. Here, the MR ratio is defined as MR-[R(H)-R(0)]/R(0). When the applied field is parallel to the current direction, the MR is negative and increases with decreasing temperature. Conversely, when the field is perpendicular to the film, the MR is positive and shows a little temperature dependence behavior. Additionally, the negative MR is essentially larger than the positive MR at temperatures below 50 K.

The most striking feature in Fig. 3 is the sign change of MR for in-plane magnetic field, independent of temperature but dependent on the relative direction of the magnetic field and the current. Since the YPtSb films studied here are non-ferromagnetic samples, we therefore speculate the sign change of MR behavior reflects the difference in current path through the film, which leads to variations in the effective localization and scattering of free carriers. For example, when the applied field is in plane, the Lorentz force is out of plane, thus there is more surface and interface scattering. In fact, the crossover of MR behavior has been previous reported in the narrow band gap semiconductor InSb films, arising from two-dimensional (2D) weak localization (WL) at the InSb/GaAs hetero interface. However, we can exclude such a contribution because, in general, the magnitude of the positive MR induced by Lorentz force is proportional to the square of the magnetic field. However, as shown in Fig. 3(b), we found that the positive MR response is approximately proportional to $H$ rather than $H^2$. Furthermore, Gorkom *et al.* [24] found that the positive MR induced by Lorentz force shows strong temperature dependence, and becomes much smaller at higher temperatures. But the positive MR shown in Fig. 3 (b) is almost temperature independence. We can therefore conclude that the positive MR induced by Lorentz force is too small to accounts for the MR variation in our sample.

More recently, Wang *et al.*[25] have proposed a new MR mechanism of the locking of the spin orientation relative to the momentum of the conduction electrons on TIs surfaces. This mechanism emphasizes that in an in-plane transport measurement configuration, the collective spin polarization of the TI surface state is aligned by the current. When the applied field is perpendicular to the current, it is parallel to spin polarization of surface current. Then a small positive MR can be observed. On the other hand, when the applied field is along the current, it is perpendicular to the spin polarization of surface current. In such a case, the angle between the spin polarization of surface current and magnetic field direction can overcome the positive MR effect and induces a negative MR behavior. This interpretation is qualitatively consistent with the observed MR behavior in the gapless YPtSb films. At this stage, however, we cannot conclude that if the YPtSb film has a topological surface state because the absent of Angle-resolved photoemission spectroscopy (ARPES) experiments. We should point out that the recent ARPES study has revealed that some half-Heuselr compounds RPtBi (R=Lu, Dy, Gd) in bulk form show a metallic surface rather than the topological surface states.[26] Nevertheless, besides a crossover of in-plane MR from negative to positive has been indentified in the high quality $Bi_2Se_3$ topological insulator thin films,[25] an unconventional linear MR and even superconductivity have been reported by Butch *et al.* in half-Heusler compound YPtBi.[27] These results do unambiguously reveal that the spin-momentum locked surface currents play a key ingredient in understanding the remarkable MR variations and topological superconductivity.

In summary, we have fabricated the gapless half-Heusler YPtSb thin films on MgO-buffered $SiO_2$/Si(001) substrates by magnetron sputtering. Investigations of the structural characteristics and electric transport properties of these films show typical a semiconducting-type half-Heusler alloy. Moreover, magneto-transport measurements demonstrate a surprisingly high Hall mobility of 450 $cm^2$/Vs at 300K, which is much higher than the value (~300 $cm^2$/Vs) obtained for the bulk YPtSb.[22] Surprisingly, the in-plane MR response to fields applied along and perpendicular to the current direction shows an opposite sign. Although our experimental results do not unambiguously justify the existence of topological surface states in the YPtSb films, we suggest that this contrasting in-plane MR response is related to the locking of the spin and current direction of the surface states. To determinate their topological surface states, an APRES measurement of ultrahigh $k$ resolution resolving both the bulk-state and surface-state contribution are needed.

Note Added. After the completion the revising of this manuscript, a related study has appeared,[28] confirming the high Hall mobility and the linear MR in gapless semiconducting polycrystalline LuPtSb sample may origin from the topological surface states.

Acknowledgements: This work was supported by National Natural Science

Foundation of China (Grant Nos. 51171207, 51021061 and 51025103) and National Basic Research Program of China (973 Programs: 2012CB619405).

**Table caption:**

Table 1. Y, Pt, and Sb mole fractions obtained by EDX form two typical films with a nominal composition of YPtSb.

| (thickness) | *Y (at.%)* | *Pt (at.%)* | *Sb (at.%)* |
|---|---|---|---|
| *0.2 μm* | 33.63 | 32.73 | 33.64 |
| *5 μm* | 33.48 | 32.17 | 34.34 |

Figure captions:

FIG. 1. (Color on line) (a) XRD spectra for simulated YPtSb lattice and fabricated YPtSb thin film. The lattice constants are found to be $a_{eq}$=6.401Å. (b) Energy dispersive X-ray spectroscopy (ED), showing the composition of YPtSb with a ratio close to 1:1:1. The inset shows the typical FE-SEM surface image of fabricated YPtSb thin film. The 's' indicates the peak of the substrate.

FIG. 2. (Color on line) Temperature-dependent zero-field electrical resistivity ρ (T) (a), Hall mobility $μ_h$ (left scale) (b) and carrier concentration n (right scale) (b) of YPtSb thin films. Inset of (a) shows the temperature-dependent magnetic-susceptibility χ(T)

FIG. 3. (Color on line) The in-plane magnetoresistance, MR=$[R(H)-R(0)]/R(0)$, where $R(H)$ is the resistance under a finite magnetic field and $R(0)$ under zero field, is shown as a function of magnetic field measured at different temperature for two distinct cases as sketched in the figures: (a) magnetic field parallel to the current and the film surface and (b) magnetic field parallel to the film surface but perpendicular to the current.

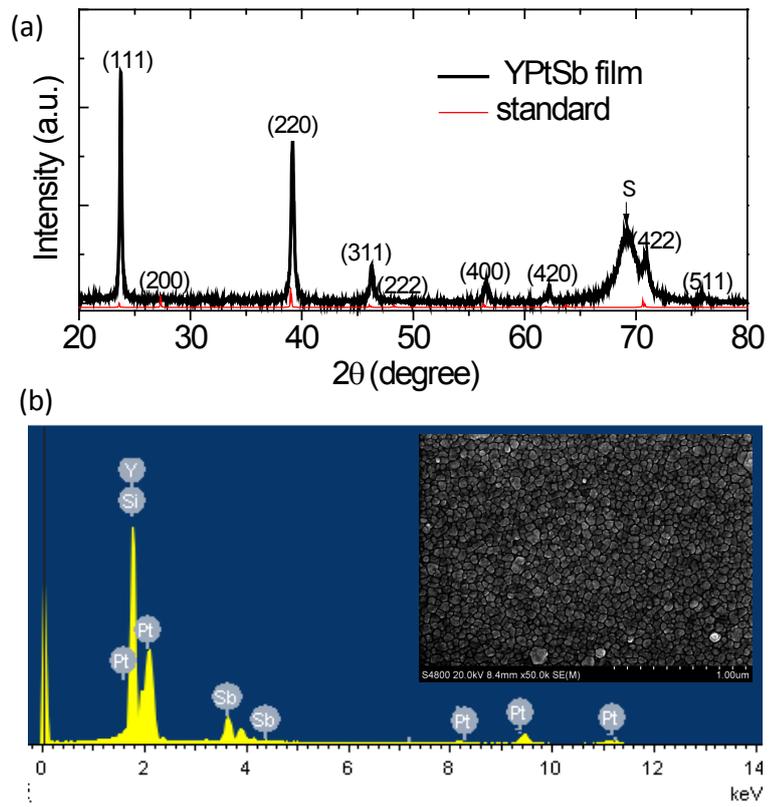

Figure 1

W. H. Wang et al., for JAP

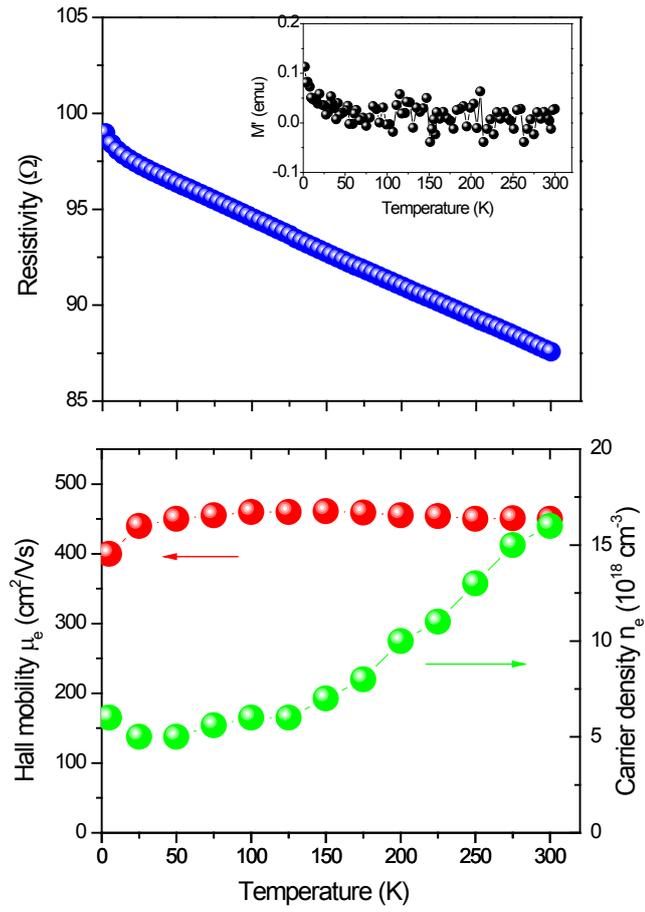

Figure 2

W. H. Wang et al., for JAP

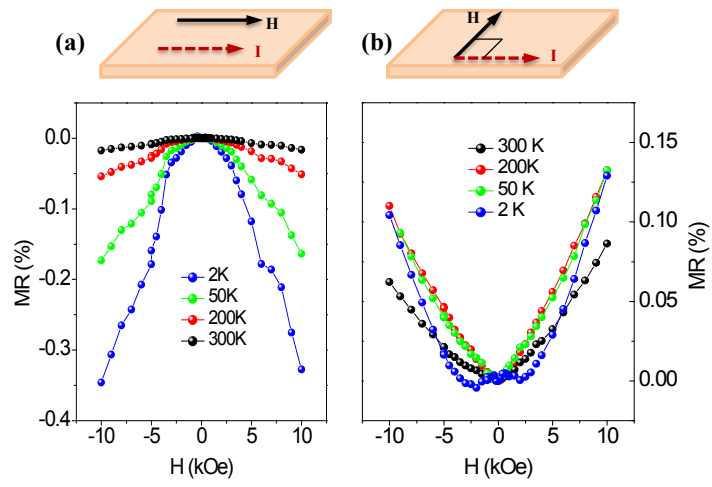

Figure 3

W. H. Wang et al., for JAP